\documentclass[10pt]{article}

\begin{document}

\title{Accelerated expansion from braneworld models with variable vacuum energy}
\author{J. Ponce de Leon\thanks{E-mail: jponce@upracd.upr.clu.edu, jpdel@astro.uwaterloo.ca}\\ Laboratory of Theoretical Physics, Department of Physics\\ 
University of Puerto Rico, P.O. Box 23343, San Juan, \\ PR 00931, USA} 
\date{April 2004}

\maketitle
\begin{abstract}

We study the acceleration of the universe as a consequence of the time evolution of the vacuum energy in cosmological models based in braneworld theories in $5D$. A variable vacuum energy may appear if the size of the extra dimension changes during the evolution of the universe. In this scenario the acceleration of the universe is related not only to the variation of the cosmological term, but also to the time evolution of $G$ and, possibly, to the variation of other fundamental ``constants" as well. This is because the expansion rate of the extra dimension
appears in different contexts, notably in expressions concerning the variation of rest mass and electric charge.
We concentrate our attention on spatially-flat, homogeneous and isotropic, brane-universes where the matter density decreases as an inverse power of the scale factor, similar (but at different rate) to the power law in FRW-universes of general relativity.
 We show that these braneworld cosmologies  are consistent with the observed accelerating universe and other observational requirements. In particular, $G$ becomes constant and $\Lambda_{(4)} \approx const \times H^2$ asymptotically in time. Another important feature is that the models contain no ``adjustable" parameters. All the quantities, even the five-dimensional ones, can be evaluated  by means of  measurements  in $4D$. We provide precise constrains on the cosmological parameters 
and demonstrate that the ``effective" equation  of state of the universe can, in principle, be determined by measurements of the deceleration parameter alone. We give an explicit  expression relating the density parameters $\Omega_{\rho}$, $\Omega_{\Lambda}$ and the deceleration parameter $q$. These results constitute concrete predictions that may help in observations for an experimental/observational test of the model.

 \end{abstract}



\newpage
\section{Introduction}

A number of recent observations suggest the possibility that our universe is speeding up its expansion \cite{Knop}-\cite{Perlmutter}. The ratio of the (baryonic and cold non-baryonic dark)  matter density to the critical density is approximately $1/3$, while the remaining  $2/3$ comes from some kind of matter whose composition is unknown and which remains unclustered on all scales where gravitational clustering of ordinary (baryonic plus dark) matter is seen. 
Measurements of the luminosity-redshift relation observed for distant supernovae of type Ia, combined with data from the cosmic microwave background radiation, suggest that this new form of matter, which is sometimes  called missing  or dark energy, should possess a large negative pressure \cite{Pelmutter2}.

The simplest candidate for the missing energy is the cosmological constant \cite{Riess}-\cite{Padmanabhan0}, which looks like an ideal fluid with negative pressure  $p_{\Lambda} = - \rho_{\Lambda}$. Another possibility is quintessence, which is a slowly-varying, spatially inhomogeneous field  \cite{Armendariz}-\cite{Caldwell1} whose effective equation of state  $p_{Q} = w_{Q}\rho_{Q}$ can vary with time and can take any value $w_{Q} > - 1$. Alternative explanations for the acceleration of the universe, beyond dark energy, include phantom energy \cite{Caldwell2}-\cite{Stefancic}, certain modifications of general relativity \cite{Turner1}-\cite{Yungui}, the gravitational leakage into extra dimensions \cite{Lue}-\cite{Deffayet2}, Chaplygin gas \cite{Gorini}-\cite{Bento} as well as Cardasian models \cite{Freese}-\cite{Gondolo}. 

In the present work, we study the acceleration of the universe as a consequence of the time evolution of the vacuum energy in cosmological models based on braneworld theories in $5D$. Our aim is to examine the interplay between these theoretical models  and observations, with some detail. 
Firstly, under very general conditions, we test whether these models  are consistent with present cosmological observations. Secondly, we derive explicit expressions among the observational quantities and obtain the specific  constraints that these models impose on the cosmological parameters.

In brane theories our four-dimensional universe is depicted as a singular hypersurface \cite{RS2},  a  $3$-brane, embedded in a $4 + d$ dimensional spacetime, or bulk. The scenario is that the matter fields are confined to the brane, while gravity propagates in the extra $d$ dimensions as well. 
The effective equations for gravity in $4D$ were obtained by Shiromizu {\it et al} \cite{Shiromizu} by using Israel's boundary conditions and imposing ${\bf Z}_2$ symmetry for the bulk spacetime, about our brane-universe {\em fixed} at some $y = y_{0}$. These equations predict five-dimensional corrections $(d = 1)$ to the usual general relativity in $4D$ and, what is specially relevant to us, they  provide explicit expressions that link the tension of the brane to the observed gravitational coupling $G$ and cosmological term $\Lambda_{(4)}$. 

Most of the studies of braneworld cosmologies in $5D$ deal with the case where the size of the fifth dimension is constant, i.e., $g_{44} = const$ \cite{Maartens}-\cite{Binetruy}. This is consistent with the assumption of time-independent vacuum energy, which in turn leads to that $\Lambda_{(4)}$ and $G$ be   constants automatically. Thus, 
if the dark energy is identified with a time-independent vacuum energy (positive cosmological constant),  then the universe will expand forever, the energy density of all kinds of matter, except for vacuum, tend to zero exponentially, and the spacetime metric approaches the de Sitter one. 

However, this simple picture is altered if the vacuum energy is allowed to vary with time.  
The crucial point here is that the introduction of a 
time-varying $g_{44}$ in braneworld theory allows the construction of a number of  cosmological 
models in $4D$ which have good  physical properties but do not admit  constant vacuum energy \cite{jpdel1}. This physical scenario cannot be ruled out {\em a priori}.
After all, the constancy of  $g_{44}$ is an external condition and not a requirement of the field equations. In this scenario $\Lambda_{(4)}$ and $G$ vary in cosmological time, which in turn influences the  expansion of the universe, through the generalized Friedmann equation.

In a recent work \cite{jpdel2} we studied the cosmological consequences of a braneworld theory, with time dependent vacuum energy, where the time variation of $G$ can be expressed as $(\dot{G}/G) = g H$ with  $|g| \leq 0.1$, in order to satisfy observational requirements.  The main results emerging from that work were that: (1) Our universe must be  spatially-flat $(k = 0)$, (2) the models with zero and negative bulk cosmological constants agree with the observed accelerating universe, while fitting simultaneously the observational data for the density and deceleration parameter, and (3) the universe must recollapse if it is embedded in an anti-de Sitter bulk, although we are nowhere near the time of recollapse. 

The question arises of whether or not these results are typical in braneworld cosmologies with variable vacuum energy. In order to elucidate this question we need to examine other models of this kind, 
based on the simplest and most general assumptions possible. In this work we use  two assumptions that are common in modern cosmology: 
\begin{enumerate}
\item The matter density decreases as an inverse power of the scale factor. 
\item The universe is spatially flat. 
\end{enumerate}

Our aim in this work is to study the cosmological  consequences of the above assumptions on the evolution of braneworld cosmologies. We will discuss the following questions: What is the general evolution of the scale factor? Does it agree with the observed accelerating universe? Is the variation of $G$ and  $\Lambda_{(4)}$ compatible with observations? What is the effective equation of state of the universe?
What is the most general relationship between density and deceleration  parameters? 
Based on our model and observational data, how to estimate and/or constrain the value of cosmological  parameters?. Although the  model is classical, we also tackle the question of whether the vacuum energy in our model can be identified with the energy of some scalar field rolling-down its potential.

\medskip

The main conclusions from our work are the following:

\medskip

The general behavior of the braneworld cosmologies under consideration can be separated into three stages. The first one is the early universe, when the quadratic correction term is dominant and the deceleration parameter is positive $(q > 0)$. As the universe expands and ages, the quadratic term decreases while the terms for matter and vacuum increase. The second stage in when the vacuum energy density and the matter energy density become equal to each other. This occurs for $\Omega_{\rho} =  1/2$.  
At this stage the effects due to vacuum become important; the matter density parameter $\Omega_{\rho}$ starts decreasing (instead of increasing) and $q$ approaches zero.  In the  third stage the universe is dominated by the vacuum energy,  which is responsible for the observed present acceleration. Since the quadratic term is now negligible, the universe can be considered as a mixture of dust and a variable cosmological ``constant" . However, the effective equation of state of the universe, for the total pressure and density,  is quintessence-like namely, $p_{t} \approx - 0.7 \rho_{t}$, which coincides with the one obtained from other models based on FRW cosmologies \cite{Efstathiou}-\cite{Steinhard}.
 
From a conceptual viewpoint, these cosmologies present a consistent picture where the current acceleration of the universe is not an isolated phenomenon, but it is intrinsically interwoven with the time variation of the cosmological parameters $G$ and $\Lambda_{(4)}$. Since the expansion rate of the extra dimension appears in different contexts such as the variation of rest mass \cite{jpdel3} and the fine-structure-constant \cite{jpdel4}, the braneworld scenario may provide us a theoretical framework  to unify all these, apparently, separated  phenomena as different consequences in $4D$ of the  time evolution of the extra dimension. This is a new step toward  understanding how the universe works.

From an observational viewpoint, based on  the positivity of the mass/energy density and the observational fact that $|g| \leq 0.1$, we provide precise constrains on the cosmological parameters.  
We show that the equation  of state of the universe can, in principle, be determined by measurements of the deceleration parameter alone. We obtain an explicit  expression between the density parameters $\Omega_{\rho}$, $\Omega_{\Lambda}$ and the deceleration parameter $q$. These results constitute concrete predictions that may help in observations for an experimental/observational test of the model.

This paper is organized as follows. In section $2$ we give a brief summary of the equations for homogeneous cosmologies in $5D$ based on braneworld theory. In section $3$ we discuss the mathematical consequences of the two assumptions mentioned above. The integration of the equations is presented in section $4$. Some of the physical consequences are explored in section $5$; we derive relationships among the observational parameters, including the age of the universe and the effective equation of state of the universe. We also show how to evaluate the five-dimensional quantities that appear in the four-dimensional equations. In sections $6$ we study the experimental/observational restrictions on the cosmological parameters. In section $7$ we present an outline of the evolution of the universe using the results of our model. Section $8$ is a summary. Finally, in the appendix we construct the scalar field potential corresponding to the vacuum energy in our model.

\section{Homogeneous cosmology in $5D$}
In order to facilitate the discussion, and set the notation, we start with a brief summary of the pertinent ideas and equations in the braneworld scenario. In this scenario our homogeneous and isotropic universe is envisioned as a singular hypersurface embedded in a five-dimensional  manifold with metric 

\begin{equation}
\label{cosmological metric}
d{\cal{S}}^2 = n^2(t,y)dt^2 - a^2(t,y)\left[\frac{dr^2}{(1 - kr^2)} + r^2(d\theta^2 + \sin^2\theta d\phi^2)\right] - \Phi^2(t, y)dy^2,
\end{equation}
where $t, r, \theta$ and $\phi$ are the usual coordinates for a spacetime with spherically symmetric spatial sections and $k = 0, +1, -1$.  The metric is a solution of the five-dimensional Einstein equations
\begin{equation}
\label{field equations in 5D}
{^{(5)}G}_{AB} = {^{(5)}R}_{AB} - \frac{1}{2} g_{AB}{^{(5)}R} = {k_{(5)}^2} {^{(5)}T_{AB}}, 
\end{equation}
where ${^{(5)}T}_{AB}$ is the five-dimensional energy-momentum tensor and $k_{(5)}$ is a constant introduced for dimensional considerations. 
 
The energy-momentum tensor on the brane $\tau_{\mu\nu}$ is separated in  two parts, 
\begin{equation}
\label{decomposition of tau}
\tau_{\mu\nu} =  \sigma g_{\mu\nu} + T_{\mu\nu},
\end{equation} 
where $\sigma$ is the tension of the brane in  $5D$, which is interpreted as the vacuum energy of the braneworld, and $T_{\mu\nu}$ represents the energy-momentum tensor of ordinary matter in $4D$. 

 There are two assumptions relating the physics in $4D$ to the geometry of the bulk. The first one is that bulk spacetime possesses   ${\bf Z_{2}}$ symmetry about the brane. As a consequence the matter in $4D$ becomes completely  determined by the geometry in $5D$ through Israel's boundary conditions. Namely, for a perfect fluid 
\begin{equation}
\label{EMT for perfect fluid}
T_{\mu\nu} = (\rho + p)u_{\mu}u_{\nu} - p g_{\mu\nu},
\end{equation}
with 
\begin{equation}
\label{equation of state for ordinary matter}
p = \gamma \rho, 
\end{equation}
the matter density is given by 
\begin{equation}
\label{density}
\rho = \frac{(- 2)}{k_{(5)}^2 (\gamma + 1)\Phi|_{brane}}\left[\frac{a'}{a} - \frac{n'}{n}\right]_{brane}.
\end{equation}
and
\begin{equation}
\label{tension of the brane}
\sigma =   \frac{(- 2)}{k_{(5)}^2 (\gamma + 1)\Phi|_{brane}}\left[(3\gamma + 2)\frac{a'}{a} + \frac{n'}{n} \right]_{brane},
\end{equation}
where a prime denotes derivative with respect to the extra variable $y$. 

The second  assumption is that the brane is embedded in an Anti-de Sitter bulk, i.e., ${^{(5)}T}_{AB}$ is taken as 
\begin{equation}
\label{AdS}
{^{(5)}T}_{AB} =  \Lambda_{(5)}g_{AB}, 
\end{equation}
where $\Lambda_{(5)} < 0$. Since ${^{(5)}T}_{\mu 4} = 0$, it follows that the energy-momentum tensor on the brane $\tau_{\mu\nu}$ is a conserved quantity, viz.,
\begin{equation}
\label{conservation on the brane}
\tau^{\nu}_{\mu; \nu} = 0. 
\end{equation}
Another important consequence of (\ref{AdS}) is that the field equations (\ref{field equations in 5D}) admit a first integral, namely, 
\begin{equation}
\label{first integral in the bulk}
\left(\frac{\dot{a}}{an}\right)^2 =  \frac{k_{(5)}^2 \Lambda_{(5)}}{6} + \left(\frac{a'}{a \Phi}\right)^2 - \frac{k}{a^2} + \frac{\cal{C}}{a^4}, 
\end{equation}
where ${\cal{C}}$ is  a constant of integration which arises from the projection of the Weyl curvature tensor of the bulk on the brane.   
Evaluating (\ref{first integral in the bulk}) at the brane, which is fixed at some $y = y_{brane} = const$, as well as  using (\ref{density}) and (\ref{tension of the brane}), we obtain the generalized Friedmann equation, viz., 
\begin{equation}
\label{generalized FRLW equation}
3\left(\frac{{\dot{a}}_{0}}{a_{0}}\right)^2  = \Lambda_{(4)}  + 8 \pi G \rho + \frac{ k_{(5)}^4}{12}\rho^2 - \frac{3 k}{a_{0}^2} + \frac{3 {\cal{C}}}{a_{0}^{4}}, 
\end{equation}
where $a_{0}(t) = a(t, y_{brane})$,  and  
\begin{equation}
\label{definition of lambda}
\Lambda_{(4)} = \frac{1}{2}k_{(5)}^2\left(\Lambda_{(5)} + \frac{ k_{(5)}^2 \sigma^2}{6}\right),
\end{equation}
\begin{equation}
\label{effective gravitational coupling}
8 \pi G =  \frac{k_{(5)}^4 \sigma}{6}.
\end{equation}
These quantities are interpreted as the  {\em net} cosmological term and gravitational coupling in $4$ dimensions, respectively. 

Equation (\ref{generalized FRLW equation}) contains two novel features; it relates  the   fundamental quantities $\Lambda_{(4)}$ and $G$ to the vacuum energy, and carries  higher-dimensional modifications to the Friedmann-Robertson-Walker (FRW) cosmological models of general relativity. Namely, local quadratic corrections via $\rho^2$, and the nonlocal corrections from the free gravitational field in the bulk, transmitted  by the dark radiation term ${\cal{C}}/{a^4}$.

Except for the condition that $n = 1$ at the brane, the generalized Friedmann  equation (\ref{generalized FRLW equation}) is valid for {\em arbitrary} $\Phi(t,y)$ and $n(t,y)$ in the bulk \cite{Binetruy}. This equation allows us to examine the evolution of the brane without using any particular solution of the five-dimensional field equations. In what follows we will omit the subscript $0$.

\section{Variable vacuum energy}

In equation (\ref{generalized FRLW equation}), $G$ and $\Lambda_{(4)}$ are usually assumed to be ``truly" constants. However, the vacuum energy density $\sigma$ does not have to be a constant. From (\ref{tension of the brane}) it follows that it depends on the details of the model. Indeed, we have recently shown \cite{jpdel1} that there are several models, with reasonable physical properties, for which a variable $\Phi$ induces a variation in the vacuum energy $\sigma$. 

 For variable vacuum energy, the conservation equations (\ref{conservation on the brane}) for a perfect fluid which satisfies (\ref{equation of state for ordinary matter}), yield 
\begin {equation}
\label{conservation equation in explicit form}
\dot{\rho} + 3\rho(\gamma + 1)\frac{\dot{a}}{a} =  - \dot{\sigma},
\end{equation}
For the case of {\em constant} $\sigma$, we recover  the familiar relationship between the matter energy density and the expansion factor $a$, viz., 
\begin{equation}
\label{familiar expansion factor}
\rho \sim \frac{1}{a^{3(\gamma + 1)}}.
\end{equation}
For a given $\sigma$ as a function of $a$, $\sigma = \sigma (a)$, we integrate  (\ref{conservation equation in explicit form}) and substitute the resulting function $\rho = \rho(a)$ into (\ref{generalized FRLW equation}), and thus obtain  the corresponding Friedmann equation.

The variation of the vacuum energy is deeply rooted in fundamental physics. The simplest microphysical model for a variable $\Lambda_{(4)}$, as well as for quintessence,  is the energy associated with a slowly evolving cosmic scalar field $\phi$ with some self-interaction potential $V(\phi)$ minimally coupled to gravity. The  
potentials  are suggested by particle physics, but in principle  $V(\phi)$ can be determined from supernova observations \cite{Huterer}-\cite{Gerke}.  

In this work, instead of constructing a field theory model for the time evolution of the vacuum energy, we adopt a phenomenological approach.  Namely, we 
assume that the energy density $\rho$ can be expressed in way similar to (\ref{familiar expansion factor}), i.e., that it decays as a power of the scale factor $a$. Specifically, 
\begin{equation}
\label{Matter density for non constant vacuum energy}
\rho = \frac{D}{a^{3(\beta +1)}},\;\;\; \beta \neq  - 1,  
\end {equation}
 where $D$ is a {\em positive} constant and $3(\beta + 1)$ is just a useful way of writing the power. This is convenient in order  to simplify the comparison with the case where $\sigma = const$, and with some previous results. With this assumption, equation (\ref{conservation equation in explicit form}) can be easily integrated as
\begin{equation}
\label{variable density}
\sigma = \sigma_{0} + \frac{D(\gamma - \beta)}{(\beta + 1)a^{3(\beta + 1)}},
\end{equation}
where $\sigma_{0}$ is a constant of integration.   
Thus, any deviation from the standard power law (\ref{familiar expansion factor}), i.e., $\beta \neq \gamma$,  yields a dynamical behavior for $\sigma$, implying that $G$ and $\Lambda_{(4)}$ are functions of  time. At this stage, we consider it premature to justify this  model on the basis  of fundamental physics, however in the appendix we show that (\ref{variable density}) can be obtained from a simple potential $V(\phi)$.  Our main purpose now is to explore whether  (\ref{variable density}) leads to an scenario where the evolution of the scale factor, as well as the simultaneous variation of $G$ and $\Lambda_{(4)}$ are consistent with cosmological observations.

With this aim we substitute (\ref{Matter density for non constant vacuum energy}) and (\ref{variable density}) into (\ref{generalized FRLW equation}) to obtain
\begin{eqnarray}
\label{Friedmann equation with non-constant sigma}
3\left(\frac{\dot{a}}{a}\right)^2 &=& \frac{k_{(5)}^2}{2}\left(\Lambda_{(5)} +  \frac{ k_{(5)}^2}{6}\sigma_{0}^2\right)\nonumber \\ &+& \frac{ k_{(5)}^{4}\sigma_{0}}{6}\left(\frac{\gamma + 1}{\beta + 1}\right) \frac{D}{a^{3(\beta + 1)}} + \frac{ k_{(5)}^4}{12}\left(\frac{\gamma + 1}{\beta + 1}\right)^2 \frac{D^2}{a^{6(\beta + 1)}}
-  \frac{3k}{a^2} + \frac{3 {\cal{C}}}{a^{4}}.
 \end{eqnarray}
This equation for $\beta = \gamma$ reduces to the generalized Friedmann equation  for constant vacuum energy. However,  for $\beta \neq \gamma$ it  generates models with varying $G$ and $\Lambda_{(4)}$. 

In order to discuss such models, it is convenient to introduce the quantities
\begin{eqnarray}
\label{useful notation}
x &=& a^{3(\beta + 1)}, \nonumber \\
A &=&  \frac{ k_{(5)}^4}{4}\left(\gamma + 1\right)^2D^2, \nonumber \\
B &=&  \frac{ k_{(5)}^4 \sigma_{0}}{2}\left(\gamma + 1\right)\left(\beta + 1\right)D, \nonumber \\
C &=& \frac{3 k_{(5)}^2}{2}\left(\beta + 1\right)^2\left(\Lambda_{(5)} + \frac{ k_{(5)}^2}{6}\sigma_{0}^{2}\right),
\end{eqnarray}
in terms of which (\ref{Friedmann equation with non-constant sigma}) becomes
\begin{equation}
\label{equation for x}
\left(\frac{dx}{dt}\right)^2 = A + Bx + Cx^2 - 9 \left(\beta + 1\right)^2 \left[k x^{(6\beta + 4)/(3\beta + 3)} - {\cal{C}}x^{(6\beta + 2)/(3\beta + 3)}\right].
\end{equation}
This equation admits exact integration, in terms of elementary functions, for a wide variety  of parameters $A$, $B$, $C$, $\beta$ and ${\cal{C}}$. In this work we assume $k = {\cal{C}} = 0$, which assures that at late times $\sigma \gg \rho$ our brane-model becomes indistinguishable from the flat FRW universes of general relativity.

At this point let us notice that,  the astrophysical data from BOOMERANG and WMAP indicate that our universe (which is assumed to be described by general relativity) is  flat $(k = 0)$ and  $\Lambda$-dominated. Besides,  the 
constant ${\cal{C}}$, which  is related to the bulk Weyl tensor and corresponds 
to an effective radiation term, is constrained to be small enough at the time of 
nucleosynthesis and it should be negligible today. Thus, setting $k = {\cal{C}} = 0$ in (\ref{equation for x}) we assure that at late times our model is compatible with observations.

\section{Spatially flat universe}
In this section we discuss the solutions  of (\ref {equation for x}) with   $k = {\cal{C}} = 0$. 
The time evolution of the brane-universe under consideration crucially depends on $C$. Namely, $x \sim \sinh\sqrt{C}t$ for $C > 0$, and $x \sim \sin\sqrt{- C}t$ for $C < 0$. Let us notice that $C$ can be written as $C = 3(\beta + 1)^2 \Lambda_{eff}$ with
\begin{equation}
\label{eff Lambda}
\Lambda_{eff}  = \frac{1}{2}k_{(5)}^2\left(\Lambda_{(5)} + \frac{ k_{(5)}^2 \sigma_{0}^{2}}{6}\right).
\end{equation}
 This quantity may take any arbitrary value by appropriately specifying the constants $\Lambda_{(5)}$ and/or $\sigma_{0}$. It plays the role of an ``effective" cosmological constant\footnote{In the case of constant vacuum energy (\ref{eff Lambda}) is indeed the cosmological constant in $4D$.} in (\ref{Friedmann equation with non-constant sigma}). If it is positive the scale factor can take arbitrary large values. Eventually, it dominates the evolution causing an exponential expansion as in the de Sitter model. If it is negative, the scale factor is bounded above and ultimately the universe recollapses.

In cosmological models with constant vacuum energy, the standard assumption is the vanishing of the {\em net} cosmological term $\Lambda_{(4)}$ defined as in (\ref{definition of lambda}). This suppresses the exponential expansion and the recollapse mentioned above. In the case of varying $\sigma$ the equivalent assumption is
\begin{equation}
\label{vanishing of the eff Lambda}
\Lambda_{eff} = 0.
\end{equation}
In order to avoid misunderstanding we should note that 
(\ref{vanishing of the eff Lambda}) does not imply $\Lambda_{(4)} = 0.$
It requires either $\Lambda_{(5)} =  - k_{(5)}^2 \sigma_{0}^2/6 =  0$ or $\Lambda_{(5)} = - k_{(5)}^2 \sigma_{0}^2/6 < 0$. 

\subsection{Bulk with $\Lambda_{(5)} = 0$}

For the first  case,  with $\sigma_{0} = 0$ in (\ref{variable density}), the scale factor is given by  
\begin{equation}
\label{Case already discussed}
a^{3(\beta + 1)} = \frac{k_{(5)}^2}{2}(\gamma + 1)D(t - t_{0}), \;\;\; \Lambda_{(5)} = 0,\;\; \sigma_{0} = 0.
\end{equation}
where $t_{0}$ is a constant of integration. The physical interpretation  of this solution has already been discussed in our previous work \cite{jpdel2}. It represents a brane-universe where $G$ and $\Lambda_{(4)}$ vary as
\begin{equation} 
\label{previous investigation}
\frac{\dot{G}}{G} \sim H, \;\;\;\Lambda_{(4)}  \sim H^2,  
\end{equation}
and the ratio $(\rho/\sigma) =  (\beta + 1)/(\gamma - \beta)$ remains constant in time, although varies with the epoch, i.e., with $\gamma$. In the remainder of this work, we will consider in detail the properties of the second case $(\Lambda_{(5)} = - k_{(5)}^2 \sigma_{0}^2/6 < 0)$.

\subsection{Anti de Sitter bulk}
In this case the expansion of the universe is governed by  
 \begin{equation}
\label{evolution for vanishing lambda eff}
a^{3(\beta + 1)} = 
\frac{D\left(\gamma +1\right)}{ 8 \sigma_{0}\left(\beta + 1\right)}\left[k_{(5)}^4 \sigma _{0}^2 \left(\beta + 1\right)^2 \left(t - t_{0}\right)^2 -  4\right], \;\;\;\Lambda_{(5)} < 0, \;\; \sigma_{0} \neq 0,
\end{equation}
where $t_{0}$ is a constant of integration. 
With this expression we find the ``deceleration" parameter $q = - \ddot{a}a/{\dot{a}}^2$ as 
\begin{equation}
\label{deceleration 4.1}
q = \frac{1 + 3 \beta}{2} + \frac{6}{\left(\beta + 1\right)k_{(5)}^4\sigma_{0}^2 (t - t_{0})^2},
\end{equation}
the Hubble parameter $H = \dot{a}/a$
\begin{equation}
\label{Hubble 4.1}
\frac{1}{H} = \frac{3(\beta + 1)}{2}(t - t_{0}) - \frac{6}{(\beta + 1)k_{(5)}^4 \sigma_{0}^2(t - t_{0})}, 
\end{equation}
the vacuum energy
\begin{equation}
\label{vacuum energy 4.1}
\frac{\sigma}{\sigma_{0}} = 1 +  \frac{8  (\gamma - \beta) }{(\gamma +1)\left[k_{(5)}^4 \sigma _{0}^2 \left(\beta + 1\right)^2 \left(t - t_{0}\right)^2 - 4\right]},
\end{equation}
and the ratio

\begin{equation}
\label{sigma over rho}
\frac{\sigma}{\rho} = \frac{\gamma - 2\beta - 1}{2(\beta + 1)} + \frac{1}{8}(\gamma + 1)(\beta + 1)k_{(5)}^4 \sigma_{0}^2 (t - t_{0})^2.
\end{equation}
In order to assure $a(t) > 0$ and $H > 0$, for all $t > 0$, we choose the constant of integration $t_{0}$ as
\begin{equation}
\label{initial t for 4.1}
t_{0} = - \frac{2}{k_{(5)}^2 \sigma_{0} (\beta + 1)}.
\end{equation}
Besides, for this choice the big bang occurs at $t = 0$. 
Now the positivity of $G$ from (\ref{effective gravitational coupling}) demands $ \sigma > 0$. On the other hand, selecting   $\gamma > \beta$, we assure  $(\sigma/\sigma_{0}) > 0$ for all values of $t$.  Consequently $\sigma _{0}$ must be  positive. 

In the remaining of this work we will study the physical properties of this solution.

\section{Cosmological parameters}

In order to compare our model (\ref{evolution for vanishing lambda eff})- (\ref{initial t for 4.1}) with observational data we need to derive relations among the observational quantities. Although the generalized Friedmann equation (\ref{generalized FRLW equation}) does not depend on the five-dimensional model, the task here is more complicated than in the FRW models because, we have to deal with the quantities $k_{(5)}$, $\sigma$. 

It is not difficult to show, using  the above formulae, that the  
 density and deceleration parameters, $\Omega_{\rho}$ and $q$, are given by
\begin{equation}
\label{density parameter}
\frac{8 \pi G \rho}{3 H^2} = \Omega_{\rho} = \frac{2(\beta + 1)[(\beta + 1) + (\gamma - \beta)(\rho/\sigma_{0})]}{(\gamma + 1)[2(\beta + 1) + (\gamma +1)(\rho/\sigma_{0})]},
\end{equation}
and
\begin{equation}
\label{deceleration parameter}
q = \frac{1 + 3\beta}{2} + \frac{3(\gamma + 1)(\beta + 1)(\rho/\sigma_{0})}{2[2(\beta + 1) + (\gamma + 1)(\rho/\sigma_{0})]}. 
\end{equation}
Thus, for any value of $\gamma$ (say dust $\gamma = 0$), if we are able to measure $\Omega_{\rho}$ and $q$, then  we obtain the present values of $(\rho/\sigma_{0})$ and $\beta$.
These can be used to evaluate the cosmological ``constant" $\Lambda_{(4)}$ in terms of the Hubble parameter. Namely, 
\begin{equation}
\label{cosm parameter}
\Lambda_{(4)} = \frac{3(\gamma - \beta) [2(\beta + 1) + (\gamma -\beta)(\rho/\sigma_{0})]}
{(\gamma + 1)[2(\beta + 1) + (\gamma + 1)(\rho/\sigma_{0})]}H^2.
\end{equation}
We note that $\Lambda_{(4)}$ here is not strictly proportional to $H^2$ because $(\rho/\sigma_{0})$ is a function of time. They become proportional only asymptotically, for $(\rho/\sigma_{0}) \rightarrow 0$. A similar situation happens with the ratio $\dot{G}/G$ where we have
\begin{equation}
\label{ratio G dot to G}
\frac{\dot{G}}{G} = - 3(\gamma - \beta)\left(\frac{\rho}{\sigma}\right)H.
\end{equation}
So, unlike in our previous study \cite{jpdel2}, in the present work $\dot{G}/G \neq const \times H$. It is important  to note that the five-dimensional quantities  $k_{(5)}$ and $\Lambda_{(5)}$ can be evaluated by means of quantities measured in $4D$ only. Specifically, for $k_{(5)}$ we have
\begin{equation}
k_{(5)}^4 = \frac{48 \pi G}{\rho}\left[1 + \frac{(\gamma - \beta)}{(\beta + 1)}\left(\frac{\rho}{\sigma_{0}}\right)\right]^{-1}\left(\frac{\rho}{\sigma_{0}}\right). 
\end{equation}
Similarly, from (\ref{eff Lambda}) and (\ref{vanishing of the eff Lambda}) we obtain $\Lambda_{(5)}$.

Let us now construct  a working expression for the age of the universe $\cal{T}$.  First  we isolate $(t - t_{0})$ from (\ref{Matter density for non constant vacuum energy}) and (\ref{evolution for vanishing lambda eff}) as
\begin{equation}
\label{t minus t zero}
(t - t_{0})^2 = \frac{4[2(\beta + 1) + (\gamma + 1)(\rho/\sigma_{0})]}{(\rho/\sigma_{0})(\gamma + 1)(\beta + 1)^2(k_{(5)}^4 \sigma_{0}^2)}.
\end{equation}
Next, we substitute this into $H^2$ from (\ref{Hubble 4.1}) and obtain
\begin{equation}
\label{characteristic time}
\frac{k_{(5)}^2 \sigma_{0}}{H} = \frac{6(\beta + 1)}{\sqrt{(\rho/\sigma_{0})(\gamma + 1)
[2(\beta + 1) + (\gamma + 1)(\rho/\sigma_{0})]}}. 
\end{equation}
This is a dimensionless quantity, which is perfectly well defined for each set $(\Omega_{\rho}, q)$. Finally, from (\ref{initial t for 4.1}), (\ref{t minus t zero}) and (\ref{characteristic time}) we obtain the desired expression, viz., 
\begin{equation}
\label{age of the universe}
{\cal{T}}H = \frac{[2(\beta + 1) + (\gamma + 1)(\rho/\sigma_{0})]}{3(\beta + 1)^2}\left\{1 - \frac{\sqrt{(\gamma + 1)(\rho/\sigma_{0})}}{\sqrt{[2(\beta + 1) + (\gamma + 1)(\rho/\sigma_{0})]}}\right\}
\end{equation}
\subsection{Parameters as functions of $q$ and $\Omega_{\rho}$}

It is useful to express the above quantities solely in terms of the observables $q$ and $\Omega_{\rho}$. To this end we use (\ref{density parameter}) and (\ref{deceleration parameter}) to eliminate $\beta$ and $\rho/\sigma_{0}$. 

From (\ref{deceleration parameter}) we obtain $\rho/\sigma_{0}$ as a function of $q$ and $\beta$. Namely, 
\begin{equation}
\label{rho over sigma zero}
\frac{\rho}{\sigma_{0}} = \frac{(\beta + 1)[2(q + 1) - 3(\beta + 1)]}{(\gamma + 1)[3(\beta + 1) - (q + 1)]}.
\end{equation}
On the other hand, from (\ref{density parameter}) we obtain $(\rho/\sigma_{0})$ as a function of $\beta$ and $\Omega_{\rho}$. From these two equations we get a quadratic equation for  $(\beta + 1)$. Namely{\footnote{Notice  that $\Omega_{\rho}$ and $q$ separate for $(\gamma + 1) = 2(\beta + 1) $, i.e., $\beta = (\gamma - 1)/2$. For this value, the density parameter is constant, viz.,  $\Omega_{\rho} = 1/2$ for all $q$.},  
\begin{equation}
\label{quadratic equation}
6(\beta + 1)^2 - 4(q + 1)(\beta + 1) + [2(\gamma + 1)(q + 1) - 3(\gamma + 1)^2 \Omega_{\rho}] = 0.
\end{equation}
In order to solve this equation, is important to  notice that the condition 
$0  < (\rho/\sigma_{0}) < \infty$ in (\ref{rho over sigma zero}) requires
\begin{equation}
\label{restriction 1}
\frac{1}{3} <  \left(\frac{\beta + 1}{q + 1}\right) < \frac{2}{3}.
\end{equation} 
Consequently, the physical solution for $\beta$, i.e. the one which  leads to a positive $\rho/\sigma_{0}$, is given by 
\begin{equation}
\label{physical solution for beta}
(\beta + 1) = \frac{q + 1}{3}\left[1 + \sqrt{1 - 3 \left(\frac{\gamma + 1}{q + 1}\right) + \frac{9}{2}\left(\frac{\gamma + 1}{q + 1}\right)^2 \Omega_{\rho}}\right].
\end{equation}
The second solution, with a negative sign in front of the root, is outside of the allowed region (\ref{restriction 1}) and leads to a negative $\rho/\sigma_{0}$. 

In order to simplify the notation, let us introduce the auxiliary function
\begin{equation}
\label{f}
f(q, \Omega_{\rho}, \gamma) = 1 + \sqrt{1 - 3 \left(\frac{\gamma + 1}{q + 1}\right) + \frac{9}{2}\left(\frac{\gamma + 1}{q + 1}\right)^2 \Omega_{\rho}},
\end{equation}
and denote
\begin{equation}
\label{xi}
\xi = \frac{q + 1}{\gamma + 1}. 
\end{equation}
With this notation, we have
\begin{eqnarray}
\label{parameters in terms of f and xi}
(\beta + 1) &=& \frac{\xi f}{3}(\gamma + 1),\nonumber \\
\frac{\rho}{\sigma} &=& \frac{\xi f (2 - f)}{3 - \xi f(2 - f)}; \;\;\;\;\;\;\frac{\rho}{\sigma_{0}} = \frac{\xi f (2 - f)}{3 (f - 1)},\nonumber \\
\Omega_{\Lambda} &=& \frac{1}{9}{(3 - \xi f)[3 - \xi (2 - f)]},\nonumber \\
{\cal{T}}H &=& \frac{1}{(\gamma + 1)(f -1)\xi}\left[1 - \sqrt{\frac{2 - f}{f}}\right].
\end{eqnarray}
In the above equations $\Omega_{\Lambda} = \Lambda_{(4)}/3H^2$ and $1 < f(q, \Omega_{\rho}, \gamma) <  2$. The positivity of $G$ requires $(\gamma - \beta) = (q + 1)(3 - \xi f)/3\xi \geq 0$, which in turn imposes an upper bound on $\xi$, namely,
\begin{equation}
\label{condition on xi}
\xi < 3.
\end{equation} 
It is interesting to note that $G$ decreases in time. Namely,  from (\ref{ratio G dot to G}) and the above expressions, we obtain 
\begin{equation}
\label{new G dot over G}
\frac{\dot{G}}{G}  = - \frac{\xi f(3 - \xi f)(2 - f)(1 + \gamma)}{3 - \xi f(2 - f)}H. 
\end{equation}
Clearly, $\dot{G}/G < 0$, as long as $(\gamma + 1) > 0$, which means that gravity was more important before than at the present time.

In addition we note  that any combination of $q$ and $\gamma$, giving the same $\xi$ in (\ref{xi}),  yields identical values for $\rho/\sigma$ and $\Omega_{\Lambda}$. Only $\beta$, the age of the universe ${\cal{T}}$ and $\dot{G}/G$ are sensible to the specific choice of $\gamma$ in  the equation of state for ordinary matter.

\subsection{The equation of state of the universe}

In order to complete our set of equations, we find  the equation of state for the {\em total} energy density $\rho_{t}$ and pressure $p_{t}$ of the universe. It can be written as 
\begin{equation}
\label{equation of state}
\frac{p_{t}}{\rho_{t}} = w,
\end{equation}
where
\begin{equation}
\label{w}
w = \beta + \frac{(\gamma + 1)(\beta + 1)(\rho/\sigma_{0})}{[2(\beta + 1) + (\gamma + 1)(\rho/\sigma_{0})]} = \frac{2q -1}{3}.
\end{equation}
The value of $w$ is not constant and differs from the equation of state $\gamma =p/\rho$ for ordinary matter on the brane. Thus, we arrive at an important result: the equation of state of the universe can be determined by measurements of the deceleration parameter alone. 

Finally, we would like to note the following relationship between the observational quantities $\Omega_{\rho}$, $\Omega_{\Lambda}$, $q$ and $\gamma$. Namely,
\begin{equation}
\label{rel bet Omegas, q and gamma}
\frac{3(\gamma + 1)^2}{2}\left[\Omega_{\rho} - \frac{2 \beta \Omega_{\Lambda}}{(\gamma - \beta)}  \right] + q(1 - \gamma) - \gamma(1 - 3 \beta) - 2 = 0.
\end{equation}
Since the overwhelming time of the evolution of the universe is spent in the matter-dominated $(p = 0)$ domain, this equation is of practical use for $\gamma = 0$ . Thus,  
\begin{equation}
\label{q}
q = 2 - \frac{3}{2}(\Omega_{\rho} + 2 \Omega_{\Lambda}).
\end{equation}
Consequently, for the equation of state of the universe we get
\begin{equation}
\label{equation of state for the whole dust universe}
w = 1 - \Omega_{\rho} - 2 \Omega_{\Lambda}.
\end{equation}
It is important to mention that the equations in this section reduce to the appropriate FRW-counterparts for $\Omega_{\Lambda} = 0$ and $\Omega_{\rho} = 1$. In particular we obtain $q = 1/2$ and $w = 0$ as in the dust-FRW cosmologies. In the next section we will use the above expressions to evaluate the parameters of the model.

\section{Physical restrictions on $(\Omega_{\rho}, q)$}

Thus, a wealth of important information about our universe can be derived from $(\Omega_{\rho}, q)$.  However, not all sets of values 
$(\Omega_{\rho}, q)$ can generate adequate  physical models. The object of this section is three-fold. Firstly, to  analyze the constraints that observation can place on them. Secondly, to  show that our model agrees with the observed accelerating universe, and fits observational data. Thirdly, to estimate the accuracy with which $\Omega_{\rho}$ and $q$ can be constrained by observations,  based on  our model.   

\subsection{The positive-energy condition}

We have already used the condition $\rho > 0$ to single out the physical solution of $\beta$. But this is not all, the positive-energy condition is a powerful tool which allows us to discriminate between what is possible or not. Here we use this condition to formulate some specific constraints on $q$ and/or $\Omega _{\rho}$.

To this end, we substitute our auxiliary function $f(q, \Omega_{\rho}, \gamma)$ into (\ref{quadratic equation}) and obtain,
\begin{equation}
\frac{2}{3}f(2 -f) = 2 \left(\frac{\gamma + 1}{q + 1}\right) - 3 \left(\frac{\gamma + 1}{q + 1}\right)^2 \Omega_{\rho}. 
\end{equation}
Then, from the positive-energy condition (\ref{restriction 1}), which in terms of $f$ becomes  $1 < f <2$, it follows that  
 \begin{equation}
\label{restriction 2}
- \frac{2}{3} < 3 \left(\frac{\gamma + 1}{q + 1}\right)^2 \Omega_{\rho} - 2\left(\frac{\gamma + 1}{q + 1}\right) < 0.
\end{equation}
This is an important expression because it  involves   observational quantities only, viz., $q$, $\Omega_{\rho}$ and $\gamma$. We now proceed to discuss its physical implications.

\subsubsection{Restrictions on $\xi = (q + 1)/(\gamma + 1)$}

We have already mentioned that for $\beta = (\gamma - 1)/2$ the density parameter is constant and equal to $1/2$, regardless of the specific value of $\gamma$ and $q$. We can easily show that the solutions that become vacuum dominated are those with  $\beta < (\gamma - 1)/2$. Indeed,  from (\ref{density parameter}) and (\ref{cosm parameter}) we get 
\begin{equation}
\Omega_{\rho} \approx \frac{\beta + 1}{\gamma + 1}, \;\;\;\;\Omega_{\Lambda} \approx \frac{\gamma - \beta }{\gamma + 1},
\end{equation}
asymptotically,  as $(\rho/\sigma_{0}) \rightarrow 0$.
In this limit, $\Omega_{\Lambda} = \Omega_{\rho}$ for $\beta = (\gamma - 1)/2$; the vacuum energy ultimately dominates only if $\beta < (\gamma - 1)/2$. For these solutions $\Omega_{\rho}$ is always positive and less that $1/2$.

The left-hand-side of (\ref{restriction 2}) provides a quadratic expression for $\xi$, which for $\Omega < 1/2$ requires either  $\xi < (3/2)[ 1 - \sqrt{1 - 2\Omega_{\rho}}]$ or $\xi > (3/2)[ 1 + \sqrt{1 - 2\Omega_{\rho}}]$, while the right-hand-side requires $\xi > 3\Omega_{\rho}/2$. 
Thus, for every given value of $\Omega_{\rho} < 1/2$,  there are two regions where the solution is consistent with $\rho 
> 0$. These are  
\begin{equation}
\label{first region}
 \left(\frac{q + 1}{\gamma + 1}\right) > \frac{3}{2}[ 1 + \sqrt{1 - 2\Omega_{\rho}}], \end{equation}
and 
\begin{equation}
\label{restriction 3}
\frac{3\Omega_{\rho}}{2} < \left(\frac{q + 1}{\gamma + 1}\right) < \frac{3}{2}[1 - \sqrt{1 - 2\Omega_{\rho}}].
\end{equation}
We now proceed to discuss the physical properties of the solution in each region.

\paragraph{Decelerated expansion:} Let us calculate the cosmological parameters in the  region controlled by (\ref{first region}). If  we take $\Omega_{\rho} = 0.1$, then the allowed values of $\xi$ are spread over a small range, viz., $ 2.842 < \xi < 2.850$. The lower limit comes from (\ref{first region}), while the upper one comes from $G > 0$, in agreement with (\ref{condition on xi}). For these values, $\Omega_{\Lambda}$ is about  $0.001$, indicating  that the evolution here is dominated by the quadratic correction term $\Omega_{\rho^2}$. The huge gravitational attraction produced by this term $(\Omega_{\rho^2} \approx 0.899)$ explains the large deceleration parameter $q \approx 1.846 + 2.846 \gamma$, which is positive for all forms of matter   including dust $(\gamma = 0)$, radiation $(\gamma = 1/3)$, stiff matter $(\gamma = 1)$ and a network of cosmic strings $(\gamma = - 1/3)$.

The same picture emerges if we use other values for $\Omega_{\rho}$ in (\ref{first region}).
Consequently, if the observational parameters  $(\Omega_{\rho}, q, \gamma)$ are tied up as in (\ref{first region}), then the solution is relevant to the description of an universe in decelerated expansion. We should mention at this point that there is plenty of observational evidence for a decelerated universe in the recent past, see e.g. \cite{Riess2}-\cite{Turner3}.  

\paragraph{Accelerated expansion:} A similar calculation,  but in the regime governed by (\ref{restriction 3}), yields a totally different picture. Namely, the universe is dominated  by the $\Lambda_{(4)}$-term and is expanding with acceleration. In view of the relevance of this type of solutions to the recent discovery of an accelerating universe, we will examine the cosmological parameters in more detail. Our aim is to see whether they fit observational data.

In  table $1$ we consider the behavior of dust-filled universes  for several values\footnote{A reliable and definitive determination of $\Omega_{\rho}$ has thus far eluded cosmologists. 
Currently, the observational data  based on dynamical techniques indicate that $\Omega_{\rho} \approx 0.1 - 0.3$ seem to be the most {\em probably} options. } of $\Omega_{\rho}$ and study how (\ref{restriction 3}) restraints  the cosmological parameters.  We use the mean value of $q$ to calculate $\beta$ and then  obtain the ``present-day" value of $(\rho/\sigma)$, $ \Omega_{\Lambda}$, the age of the universe $\cal{T}$, as well as other relevant cosmological parameters,  such as  $w = p_{t}/\rho_{t}$ and $g = \dot{G}/HG$. 

\begin{center}
\begin{tabular}{|c|c|c|c|c|c|c|c|} \hline
\multicolumn{8}{|c|}{\bf Table 1: Determining the cosmological parameters from $\Omega_{\rho}$}\\ \hline
 \multicolumn {1}{|c|}{$\Omega_{\rho}$} & $q$& $\beta$ & 
$(\rho/\sigma)$ & $ \Omega_{\Lambda}$ & {$H {\cal{T}}$} & $w$ &
 \multicolumn{1}{|c|}{$|g|$} \\ \hline\hline
                    $0.3$ & $- 0.499 \pm 0.051$& $- 0.729$   &$0.113$& $0.683$&$1.659$& $- 0.666$&$0.248$   \\ \hline
$0.2$ &$- 0.681 \pm 0.019$ & $- 0.823$   & $0.063$& $0.794$&$2.600$ &$- 0.787$&$0.156$  \\ \hline
$0.1$ &$- 0.846 \pm 0.004$ & $- 0.913$  & $0.027$ &$0.899$&$5.379$ &$- 0.897$&$0.073$  \\ \hline
 \end{tabular}
\end{center}

Table $1$ shows that the universe is expanding with acceleration, in agreement  with  modern observations. The 
acceleration is driven by the repulsive effect of the ``dark energy" 
associated with  $\Lambda_{(4)}$, which clearly dominates the evolution here. For $\Omega_{\rho} \approx 0.3$  
the quadratic correction is ``negligible", $\Omega_{\rho^2} \approx 0.02$,   and  decreases  as the universe gets older.

Thus, at late times $\Omega_{\rho} + \Omega_{\Lambda} \approx 1$. In other words, the resulting cosmological model is a combination  of ordinary matter with  an  unknown form of matter which is effectively described by a cosmological term which varies with time. The latter satisfies the equation of state $p_{\Lambda} = - \rho_{\Lambda}$, with  positive  energy density, and dominates over the regular matter. As a consequence, at the present stage, the evolution of the universe is driven by an effective equation of state  which is  ``quintessence-like", in the sense that $(- 1 < w < 0)$.

We note that, knowing the value of the matter density $\rho$, the vacuum energy density can be obtained from the fourth column in table $1$. We can also calculate the constant $\sigma_{0}$, or $\rho/\sigma_{0}$, from (\ref{parameters in terms of f and xi}). This allows us to obtain $k_{(5)}$ from 
(\ref{characteristic time}) as well as  the value of the cosmological constant in $5D$, $\Lambda_{(5)} = - k_{(5)}^2\sigma_{0}^2/6$, which follows from  (\ref{eff Lambda}) and (\ref{vanishing of the eff Lambda}).

\subsubsection{Restriction on $\Omega_{\rho}$} 

Reciprocally, (\ref{restriction 2}) can be used to restrict the density parameter by measurements of $q$. Namely, for every given value of $q$, there is a range of possible values for $\Omega_{\rho}$, viz., 
\begin{equation}
\label{restriction 4}
\frac{2}{3}\left(\frac{q + 1}{\gamma + 1}\right)\left[1 - \frac{1}{3}\left(\frac{q + 1}{\gamma + 1}\right) \right] < \Omega_{\rho} < \frac{2}{3}\left(\frac{q + 1}{\gamma + 1}\right).
\end{equation}
If we take the deceleration parameter as in the usual dust FRW cosmologies, i.e., $q = + 0.5$, then $1/2 < \Omega < 1$. However, recent measurements indicate that, the current universe is speeding up its expansion with an acceleration parameter which is roughly $q = - 0.5 \pm 0.2$.  
If we fix $q$ today, then the adequate values of $\Omega_{\rho}$ are spread over a small range, which decreases as the universe ages. 
In table $2$ we use the mean value of $\Omega_{\rho}$ to calculate  the cosmological parameters, corresponding to various applicable values of  $q$, for  universes in the dust-dominated era $(\gamma = 0)$. 
The density parameter $\Omega_{\rho}$ can be determined from (\ref{restriction 4}) with an accuracy which is up to $86\%-95\%$.

\begin{center}
\begin{tabular}{|c|c|c|c|c|c|c|c|} \hline
\multicolumn{8}{|c|}{\bf Table 2: Determining the cosmological parameters from $q$}\\ \hline
 \multicolumn {1}{|c|}{$q$} & $\Omega_{\rho}$& $\beta$ & 
$(\rho/\sigma)$ & $ \Omega_{\Lambda}$ & {$H {\cal{T}}$} &$w$&

\multicolumn{1}{|c|}{$|g|$} \\ \hline\hline
                    $- 0.3$ & $0.412 \pm 0.054$& $- 0.602$ & $0.133$& $0.561$&$1.183$&$- 0.533$& $0.240$   \\ \hline
$- 0.5$ &$0.305 \pm 0.028$ & $- 0.717$   & $0.093$& $0.681$&$1.657$ &$- 0.666$&$0.200$  \\ \hline
 $- 0.7$ &$0.190 \pm 0.010$ & $- 0.829$ & $0.052$ &$0.805$&$2.761$& $- 0.800$&$0.131$  \\ \hline
\end{tabular}
\end{center}
It is important to note that the possibly values of $\Omega_{\rho}$ $(\Omega \approx 0.1 - 0.3)$ and $q$ ($q = -0.5 \pm 0.2$) contain an error, or uncertainty, of  about $50\%$ and $40\%$, respectively. This is a clear evidence of the difficulties involved in the  determination of $\Omega_{\rho}$ and $q$ from observations. We can  use our model to narrow down these uncertainties and get  $\Omega_{\rho}$ and $q$ with a better accuracy. 

First, from table $1$ we see that for $\Omega \approx 0.1 - 0.3$ the acceleration parameter varies in the range $- 0.85 < q < - 0.448$. Then, from the last row in table $2$ we see that large negative values for $q$, bellow  $- 0.7$,  are possible because we are considering density parameters less than $\Omega_{\rho} = 0.18$. Based upon the observational limit on $q$, we can set  a lower bound to $\Omega_{\rho}$. Consequently, 
\begin{equation}
\label{lower bound for density}
    \Omega_{\rho} \approx 0.18 - 0.3.
\end{equation}
Second, from table $2$ we see that  that $q = -0.5 \pm 0.2$ requires  $\Omega_{\rho } \approx 0.18 - 0.446$. Then, from the first row of table $1$ we see that large values for $\Omega_{\rho}$, above  $0.3$,  result from considering $q$ greater  than $\approx - 0.448$. Again,  based upon the observational limits on the density parameter,   we can set $q = - 0.448 $ as the upper limit of $q$. Thus, 
\begin{equation}
\label{first determination of q}
q = - 0.574 \pm 0.126.
\end{equation}
The conclusion is that our model allows to significantly reduce the uncertainty from $50 \%$ to $25\%$ for  $\Omega_{\rho}$,  and  from $40\%$ to $22\%$ for $q$. We emphasize, however, $\Omega_{\rho}$ and $q$  are not completely independent: fixing $\Omega_{\rho}$ today also fixes $q$ with a $90-99\%$ confidence. 

\subsection{Constraints from the variation of $G$}

We now proceed to show that our model is compatible with the observational constraints on the variation 
of $G$. To this end, we use (\ref{new G dot over G}) and denote 
\begin{equation}
g = - \frac{\xi f(3 - \xi f)(2 - f)(1 + \gamma)}{3 - \xi f(2 - f)},
\end{equation}
so that $\dot{G}/G = g H$. In astronomical and geophysical experiments, measurements of possible variations 
of $G$ with cosmic time provide experimental bounds on $\dot{G}/G$, which can be summarized as\footnote{A comprehensive and updated discussion  of the various experimental and 
observational constraints on the value of $g$ (as well as on the variation 
of other fundamental ``constants" of nature) has recently been provided 
by Uzan \cite{Uzan}} 
\begin{equation}
\label{observational constraints on g}
 |g| \leq 0.1.
\end{equation}
Clearly, the positivity of energy is not enough to guarantee the fulfillment of this condition. This is illustrated in tables $1$ and $2$. Thus, (\ref{observational constraints on g}) imposes observational constraints on $(\Omega_{\rho}, q)$.

For example, if we take $\Omega_{\rho} = 0.3$ and $\gamma = 0$,  then  (\ref{observational constraints on g}) requires $- 0.571 < q < - 0.529$. On the other hand, (\ref{restriction 3}) requires $- 0.55 < q < - 0.448$. As a consequence, the {\em physical} range of $q$ reduces to $- 0.55 < q < - 0.529$, or $q = - 0.539 \pm 0.011$. Therefore, for a fixed $\Omega_{\rho}$, the best lower bound on $q$ is provided by the positive-energy condition (\ref{restriction 3}), while the best upper bound is provided by (\ref{observational constraints on g}).

Table $3$ shows how the observational evidence $|g| < 0.1$ allows to increase the precision in the determination of $q$ to be more than $98\%$. Again, we use the mean value of $q$ to calculate the rest of the parameters. We note that smaller $\Omega_{\rho}$ means that the universe is older and $\beta$ is closer to $- 1$ today. We also note that 
\begin{equation}
\label{rel between beta and Omega Lambda}
w \approx - \Omega_{\Lambda},
\end{equation}
with an error of less than $1\%$, which is consistent with (\ref{equation of state for the whole dust universe}) for an ``old" $(\Omega_{\rho^2} \approx 0)$ universe.

\begin{center}
\begin{tabular}{|c|c|c|c|c|c|c|c|c|} \hline
\multicolumn{8}{|c|}{\bf Table 3: Cosmological parameters for various $\Omega_{\rho}$ fitting   $|g| < 0.1$}\\ \hline
 \multicolumn {1}{|c|}{$\Omega_{\rho}$} & $q$ & $\beta$ & 
$(\rho/\sigma)$ & $ \Omega_{\Lambda}$ & {$H {\cal{T}}$} &$w$&
 \multicolumn{1}{|c|}{$|g|$
} \\ \hline\hline
                     
$0.3$ & $ - 0.539 \pm 0.011$& $- 0.705$   &$0.024$& $0.696$&$1.875$ & $- 0.693$&$0.052$  \\ \hline
$0.2$ &$- 0.694 \pm 0.006$ & $- 0.806$   & $0.020$& $0.798$&$2.796$ & $- 0.796$&$0.048$  \\ \hline
$0.1$ &$- 0.847 \pm 0.003$ & $- 0.909$  & $0.020$ &$0.899$&$5.436$ & $- 0.898$&$0.054$  \\ \hline
 \end{tabular}
\end{center}
In the same way, for every given $q$  there is a range of values for $\Omega_{\rho}$ that  satisfy the inequality (\ref{observational constraints on g}). As an example consider $q = -0.3$ and $\gamma = 0$. From (\ref{observational constraints on g}) we get $0.441 < \Omega_{\rho} < 0.5$, while from (\ref{restriction 3}) we have $0.358 < \Omega_{\rho} < 0.466$. Therefore, the physical range reduces to $0.441 < \Omega_{\rho} < 0.466$. Consequently, for a fixed $q$ the upper limit of $\Omega_{\rho}$ is imposed by $\rho >0$, while the lower limit by $|g|< 0.1$.

In table $4$ we use $\rho > 0$ and $|g| < 0.1$ to constraint $\Omega_{\rho}$ for  different values of $q$. Once more, for a given $q$ the values of $\Omega_{\rho}$ are spread over a very small range, so the confidence interval  for the density parameter is up to $97\% - 98\%$. 

In order to put the above results in perspective, let us keep in mind that in FRW cosmologies $q$ and $\Omega_{\rho}$ are connected through the relation  $q = \Omega_{\rho}(1 + \gamma)/2$. Which means that $\Omega_{\rho}$ (or $q$) can be determined by measurements of $q$ (or $\Omega_{\rho}$) alone - in principle nothing else is needed. Unlike this, in our model  
  $q$ and $\Omega_{\rho}$ are, from a mathematical point of view,  {\em independent} from each other. However, from a physical point of view, they are not autonomous. The positive-energy condition creates a strong linkage between them, namely (\ref{restriction 2}), which is further strengthen by the observational evidence that $|g| < 0.1$.
What is amazing here is that these physical restrictions are strong enough as to allow the determination of one of these quantities by measurements of the other one, with a high degree of precision which today is close to $100\%$.

\begin{center}
\begin{tabular}{|c|c|c|c|c|c|c|c|} \hline
\multicolumn{8}{|c|}{\bf Table 4: Cosmological parameters for various $q$ fitting $|g| < 0.1$} \\ \hline
 \multicolumn {1}{|c|}{$q$} & $\Omega_{\rho}$& $\beta$ & 
$(\rho/\sigma)$ & $ \Omega_{\Lambda}$ & {$H {\cal{T}}$} &$w$&

\multicolumn{1}{|c|}{$|g|$} \\ \hline\hline
                    $- 0.3$ & $0.454 \pm 0.013$& $- 0.547$ & $0.028$& $0.540$&$1.252$&$- 0.533$& $0.046$   \\ \hline
$- 0.5$ &$0.326 \pm 0.007$ & $- 0.678$   & $0.022$& $0.670$&$1.743$ &$- 0.666$&$0.046$  \\ \hline
 $- 0.7$ &$0.196 \pm 0.004$ & $- 0.811$ & $0.020$ &$0.802$&$2.847$ &$- 0.800$&$0.050$  \\ \hline
\end{tabular}
\end{center}

To finish this section we would like to use the above results to ``improve" the limits discussed in (\ref{lower bound for density}) and (\ref{first determination of q}). 
From table $3$ we see that the highest $q$, corresponding to $\Omega_{\rho} = 0.3$, which satisfies  the condition $|g| \leq 0.1$ is $q \approx - 0.528$, not\footnote{The rest of the values of  $q \in (- 0.528, - 0.448)$, allowed by $\rho > 0$, in table $1$ for $\Omega_{\rho} = 0.3$,  are excluded because they  do {\em not} comply with (\ref{observational constraints on g}).} $- 0.448$ as in (\ref{first determination of q}). Since the observational data suggest that $ 0.3 \leq \Omega \leq 0.1$, it follows that $q = - 0.528$ is the upper limit for the acceleration parameter. Thus, 
\begin{equation}
\label{second determination of q}
q = - 0.614 \pm 0.086.
\end{equation} 
Similarly for the density parameter. From table $4$ we see that the lowest value of $\Omega_{\rho}$, for $q = - 0.7$,  which satisfies (\ref{observational constraints on g}) is $\Omega_{\rho } = 0.192$. The models for $q = - 0.7$ with $\Omega_{\rho} \in(0.180, 0.192)$ are excluded because they do not comply with $|g| \leq 0.1$. Consequently, from $ - 0.7 \leq q \leq - 0.3$ we get
\begin{equation}
\label{second lower bound for density}
0.192 \leq \Omega_{\rho} \leq 0.3.
\end{equation}
Consequently, our model together with the observational fact that $|g| < 0.1$,  allows us to confine the whole range of possible values for the parameters $q$ and $\Omega_{\rho}$ with and accuracy of about $86\%$ and $78\%$, respectively. 

\section{Past, present and future evolution}

The aim of this section is to outline the evolution of the universe according to our solution.

\subsection{The past}
In braneworld models, the early universe is dominated by the quadratic correction term. Accordingly, for $t \approx 0$,  in the ``very" early universe, our solution (\ref{evolution for vanishing lambda eff})  simplifies to 
\begin{equation}
\label{a for the simplest model}
a(t) \approx \left[\frac{k_{(5)}^2}{2}\left(\gamma + 1\right)D t\right]^{1/3(\beta + 1)}.
\end{equation}
Thus, 
\begin{equation}
\label{q for the simplest model}
q \approx 2 + 3\beta, \;\;\; H \approx \frac{1}{3(\beta + 1)} \frac{1}{t}.
\end{equation}
The matter energy density and the vacuum energy density are given by
\begin{equation}
\label{vacuum energy in the simplest model}
\rho \approx \frac{2}{(\gamma + 1)k_{(5)}^2 t}, \;\;\; \sigma \approx \frac{2 \left(\gamma - \beta \right)}{\left(\beta + 1\right)\left(\gamma + 1\right)k_{(5)}^2 t}.
\end{equation}
Consequently, the cosmological  term $\Lambda_{(4)}$ and gravitational coupling $G$ vary as 
\begin{equation}
\label{condition on gamma minus beta}
\Lambda_{(4)} \approx \frac{3 (\gamma - \beta)^2}{(\gamma + 1)^2}H^2, \;\;\; 8\pi G \approx  \frac{ k_{(5)}^2 (\gamma - \beta)}{\gamma +1} H.
\end{equation}
In addition, 
\begin{equation}
\label{Omega rho for the early universe}
\Omega_{\rho} = \frac{2(\beta + 1)(\gamma - \beta)}{(\gamma + 1)^2}.
\end{equation}
We note that in this limit, the solution is indistinguishable from the one for $\Lambda_{(5)} =  \sigma_{0} = 0$ presented in (\ref{Case already discussed}) and (\ref{previous investigation}). This is physically reasonable because at this stage the evolution is dominated by $\rho^2$, while the effects of the vacuum energy  become important later. As an illustration, let us take a ``small" $\Omega_{\rho}$ (say $\Omega_{\rho} = 0.01$). Then, from (\ref{Omega rho for the early universe}) it follows that $\beta \approx \gamma - \Omega_{\rho}(1 + \gamma)/2$. Consequently, $\rho/\sigma = (\beta + 1)/(\gamma - \beta) \approx 2/\Omega_{\rho} \approx 200$. 

For $\Omega_{\rho} = 1/2$, it follows from (\ref{Omega rho for the early universe})  that $\beta = (\gamma -1)/2$. Consequently,  the scale factor matches the one in  FRW models, viz.,
\begin{equation}
\label{FRW1}
a(t) \sim t^{{2}/{3(\gamma + 1)}}.
\end{equation}
Hence, 
the deceleration parameter (\ref{q for the simplest model}) 
reduces to $q \approx (1 + 3 \gamma)/2$ which is the usual one 
in FRW cosmologies. However,  
in this limit neither the gravitational coupling is constant, nor 
the cosmological term becomes zero. Instead we have
\begin{equation}
\label{FRW2}
8 \pi G = \frac{k_{(5)}^2}{2}H, \;\;\; \Lambda_{(4)} = 
\frac{3}{4}H^2, \;\;\; H = \frac{2}{3(\gamma + 1)}\frac{1}{t},
\end{equation}
where $H$ is the Hubble parameter in FRW models. We note that this solution was also found in our recent work \cite{jpdel2}. It describes the stage of the evolution when  there is a perfect balance between ordinary matter and vacuum energy, namely $\rho = \sigma$. At this stage there is no difference between the equation of state for ordinary matter and the one for the whole universe because $w = \gamma$.

Throughout the early evolution, $\rho > \sigma$, the parameters $q$, $\gamma$ and $\Omega_{\rho}$ satisfy (\ref{first region}). The expansion $(H > 0)$ is slowing down $(q > 0)$ due to the huge gravitational attraction produced by the $\rho^2$-term in the generalized Friedmann equation. 
As the universe expands; $\Omega_{\rho^2}$ and (consequently) $q$ decrease.  At the same time $\Omega_{\rho}$ and  $\Omega_{\Lambda}$ increase, such that $\Omega_{\rho^2} + \Omega_{\rho} + \Omega_{\Lambda} = 1$.
However, $\Omega_{\rho}$ is bounded above\footnote{The models for which $ 1/2 < \Omega_{\rho} < 1$ never become vacuum dominated. These are of no interest for us here.}, while $\Omega_{\Lambda}$ is not. The upper limit $\Omega_{\rho} = 1/2$ corresponds to $\rho = \sigma$. The epoch of (decelerated) expansion does not finish here; it continues until $\Omega_{\rho} \approx \Omega_{\Lambda}$, which denotes the moment when the universe becomes vacuum dominated\footnote{We note that $\sigma > \rho$, does not necessarily imply $\Omega_{\Lambda} > \Omega_{\rho}$. This is a consequence of the quadratic correction term in the Friedmann equation.} In this phase of evolution the equation of state of the universe becomes quintessence-like $(w < 0)$.

\begin{center}
\begin{tabular}{|c|c|c|c|c|c|c|c|} \hline
\multicolumn{8}{|c|}{\bf Table 5: Outline of the evolution of the universe}\\ \hline
 \multicolumn {1}{|c|}{$\Omega_{\rho}$} & $\xi$&$q$& $\beta$ & 
$(\rho/\sigma)$ & $ \Omega_{\Lambda}$ & {$(1 + \gamma)H {\cal{T}}$} & 
 \multicolumn{1}{|c|}{$w$} \\ \hline\hline
$0.01$ &$2.985 \pm 0.000$ & $1.985$  & $- 0.363 \times 10^{- 7}$ &$198.00$&$0.363 \times 10^{- 9}$ &$0.334$&$0.990$\\ \hline
$0.1$ &$2.846 \pm 0.004$ & $1.846$  & $- 0.015$ &$17.973$&$0.001$ &$0.345$&$0.897$ \\ \hline
$0.3$ &$2.499 \pm 0.050$ & $1.499$   & $ - 0.063$& $4.553$&$0.017$ &$0.378$&$0.666$ \\ \hline
$0.48$ &$2.04 \pm 0.24$ & $1.04$  & $- 0.170$ &$1.833$&$0.080$ &$0.446$&$0.360$  \\ \hline
$0.490$ &$1.988 \pm 0.276$ & $0.988$  & $- 0.191$ &$1.705$&$0.092$ &$0.458$&$0.325$  \\ \hline
$0.499$ &$1.909 \pm 0.342$ & $0.909$  & $- 0.230$ &$1.550$&$0.114$ &$0.479$&$0.273$ \\ \hline
$0.5$ &$1.5 \pm 0.75$ & $0.5$  & $- 0.5$ &$1$&$0.25$ &$0.667$&$0$ \\ \hline
$0.499$ &$1.091\pm 0.342$ & $0.091$  & $- 0.502$ &$0.458$&$0.387$ &$0.797$&$- 0.273$ \\ \hline
$0.490$ &$1.013\pm 0.274$ & $0.013$  & $- 0.516$ &$0.378$&$0.417$ &$0.846$&$- 0.325$  \\ \hline
$0.48$ &$0.960 \pm 0.240$ & $- 0.040$  & $- 0.530$ &$0.333$&$0.440$ &$0.886$&$- 0.360$  \\ \hline
$0.3$ &$0.461 \pm 0.011$ & $- 0.539$  & $- 0.705$ &$0.024$&$0.696$ &$1.875$&$- 0.693$  \\ \hline

 \end{tabular}
\end{center}
The quantities  $\rho/\sigma$ and $\Omega_{\Lambda}$ do not depend on the specific value of $\gamma$. Unlike this, $q$, $\beta$, $w$ and ${\cal{T}}$ do depend on it. In table $5$ we present their values calculated for $\gamma = 0$. For any other $\gamma$ the corresponding parameters $q_{\gamma}$, $\beta_{\gamma}$ and $w_{\gamma}$
are given by 
\begin{equation}
\label{q and w as functions of gamma}
q_{\gamma} = q(1 + \gamma) + \gamma, \;\;\;\beta_{\gamma} = \beta(1 + \gamma) + \gamma,\;\;\; w_{\gamma} = w(1 + \gamma) + \gamma.
\end{equation}

 \subsection{The present}

Current dynamical mass measurements suggest that the matter content of the universe adds up to $30 \%$ of the critical density\footnote{Radiation $0.005\%$,  ordinary luminous baryonic matter $0.5 \%$, ordinary non-luminous baryonic matter $3.5 \%$ and exotic (non-baryonic) dark matter ``observed" through their gravitational effects $26 \%$.}.  
Correspondingly, if we take $\Omega_{\rho} = 0.3$, then
the universe is speeding up its expansion with  $q = - 0.539 \pm 0.011$ (see table $3$), which is consistent with recent observations. 

Although the accelerated expansion of the universe, at the current epoch, is driven by the dominating dark energy associated with $\Lambda_{(4)}$, the  dynamical evolution is established by the {\em total} matter content, which includes regular (cold dark  plus baryonic) matter as well as the dark energy component. The effective equation of state of the universe (\ref{w}) for $q = - 0.539 \pm 0.011$ is restricted to be 
\begin{equation}
\label{specific number for w}
 - 0.7 < \left(\frac{p_{t}}{\rho_{t}}\right) < - 0.685.
\end{equation}
Going aside for a moment, let us to note that in our model the equation of state of the universe, $p_{t} = w \rho_{t}$, is similar to the one derived for quintessence in the framework of  flat FRW cosmologies \cite{Efstathiou}. In this framework, observations from  SNe Ia and CMB indicate that the equation of state for quintessence $w_{Q} = p_{Q}/\rho_{Q}$ has an upper limit $w_{Q} \approx - 0.6$, which is close to the lower limit $w_{Q} \approx - 0.7$ allowed for  quintessence tracker fields \cite{Steinhard}. 
Since the dynamics of spacetime is governed by the effective equation of state $p_{t} = w \rho_{t}$ (not only by the vacuum energy), the fact that  $w_{Q} \approx w$ is a clear evidence of how little is the effect of ordinary matter on the present evolution of the universe. 

For practical reasons, it is convenient to express the solution in terms of the quantity $z$, the redshift of the light emitted at time $t$, which is the directly observed quantity used in research on distant objects. The connection between the scale and the redshift parameter is given by 
\begin{equation}
\label{definition of z}
a(t) = \frac{a(\bar{t})}{1 + z},
\end{equation}
where $\bar{t}$ is the present time, i.e., $\bar{t} = {\cal{T}}$,  so that $z = 0$ now.  From (\ref{evolution for vanishing lambda eff}) and (\ref{initial t for 4.1}) we get
\begin{equation}
\tau^2 + 4\tau = \frac{4 \bar{\eta}}{(1 + z)^{3(\beta + 1)}},\;\;\; \tau \equiv k_{(5)}^2\sigma_{0}(\beta + 1)t,  
\end{equation}
where for convenience we have introduced the constant $\bar{\eta}$ as 
 \begin{equation}
4 \bar{\eta} \equiv \bar{\tau}^2 + 4 \bar{\tau}.
\end{equation}
Employing the assumption $\gamma = 0$ today, and (\ref{definition of z}), the cosmological parameters in terms of $z$ are
\begin{eqnarray}
\label{cosmological parameters  as a function of z}
\xi(z) &=& q(z) + 1 = \frac{3(\beta + 1)}{2}\left(1 + \frac{(1 + z)^{3(1 + \beta)}}{[\bar{\eta} + (1 + z)^{3(1 + \beta)}]}\right), \nonumber \\
8 \pi G(z) &=& \frac{k_{(5)}^4 \sigma_{0}}{6}\left[ 1 + \frac{(- \beta)}{\beta + 1}\left(\frac{\bar{\rho}}{\sigma_{0}}\right)(1 + z)^{3(1 + \beta)}\right], \nonumber \\
\Omega_{\Lambda}(z) &=& \frac{- \beta [2(\beta + 1) - \beta (\bar{\rho}/\sigma_{0})(1 + z)^{3(1 + \beta)}]}{[2(\beta + 1) + (1 + z)^{3(1 + \beta)}]}, \nonumber \\
\frac{\Omega_{\rho}(z)}{\bar{\Omega}_{\rho}} &=& \frac{(\bar{\eta} + 1)G(z)}{[\bar{\eta} + (1 + z)^{3(1 + \beta)}]\bar{G}},
\end{eqnarray}
where $\bar{\rho}$ , $\bar{G}$ and $\bar{\Omega}_{\rho}$ represent  the present values  of ${\rho}$ , ${G}$ and ${\Omega}_{\rho}$, respectively. 
Thus, in order to be able to {\em predict} the above parameters, at any $z$, in terms of their present values, we need to know the constants $k_{(5)}$, $\beta$ and $\sigma_{0}$.
\subsubsection{Evaluation of the constants}

In order to evaluate the various constants as well as the vacuum energy and the cosmological ``constant" today, let us take $\bar{\Omega}_{\rho}$, $\bar{q}$ and $\beta$ from the last row in table $5$. For these values, our universe is almost three times older than  the dust FRW universe, for which $H{\cal{T}}_{FRW} = 2/3$. Besides, the vacuum energy density is much bigger that the matter energy density, namely,
\begin{equation}
\bar{\sigma} \approx 41.67\bar{\rho},  \;\;\;\; \bar{\rho} = 3.581\times 10^{-2}\left(\frac{{\bar{H}}^2}{\bar{G}}\right).
\end{equation}
The  five-dimensional quantities $k_{(5)}$ and $\Lambda_{(5)}$ can be evaluated as 
\begin{equation}
\label{5D quantities}
k_{(5)}^2 = 0.363 \left(\frac{\bar{H}}{\bar{\rho}}\right) =  10.146 \left(\frac{\bar{G}}{\bar{H}}\right), \;\;\; \Lambda_{(5)} = -5.390 \times 10^{2} (\bar{H} \bar{\rho}) =  - 19.303 \left(\frac{{\bar{H}}^3}{\bar{G}}\right).
\end{equation}
If we take $\bar{G} = 6.67\times 10^{- 8} cm^3 gr^{- 1} s^{-2}$ and $\bar{H} \approx 0.7 \times 10^{- 10} yr^{-1} = 2.22 \times 10^{- 18} s^{- 1}$, then 
\begin{equation}
\bar{\rho} = 2.65 \times 10^{- 30} gr cm^{-3} = 1.14 \times 10^{- 47}GeV^4.
\end{equation}
The universe is  around $26.8 Byr$ old and, and  the rest of the parameters are as follows
\begin{eqnarray}
\label{numerical values}
{\bar{\Lambda}}_{(4)} &=& 1.03 \times 10^{- 35} s^{-2} = 4.46\times 10^{- 84} GeV^{ 2},\nonumber \\
{\bar{\rho}}_{\Lambda} &=& 6.14 \times 10^{-30} gr cm^{- 3} = 2.65 \times 10^{-47}GeV^4, \nonumber \\
\bar{\sigma} &=& 1.08 \times 10^{- 28} gr cm^{- 3} = 4.66\times 10^{- 46} GeV^{4}, \nonumber \\
\sigma_{0} &=& 1.02 \times 10^{- 28} gr cm^{- 3} = 4.40\times 10^{- 46} GeV^{4}, \nonumber \\
k_{(5)}^2 &=& 3.05 \times 10^{11} cm^3 gr^{-1} s^{-1} = 4.66 \times 10^{4} GeV^{- 3}\nonumber \\ 
\Lambda_{(5)} &=& - 3.17 \times 10^{- 45}cm^{- 3}gr s^{-1} = - 0.90 \times 10^{- 86} GeV^5, \nonumber \\
k_{(5)}^2 \Lambda_{(5)} &=& - 9.67 \times 10^{- 34} s^{- 2} = - 4.19 \times 10^{- 82} GeV^2.
\end{eqnarray}
 In table $6$ we illustrate how a modest change of $q$, within a range of allowed values, influences some of the parameters in the solution. 
\begin{center}
\begin{tabular}{|c|c|c|c|c|c|c|} \hline
\multicolumn{7}{|c|}{\bf Table 6: Dust universe for $\Omega_{\rho} = 0.3$, and various $q$}\\ \hline
 \multicolumn {1}{|c|}{$q$} & $\beta$ & 
$(\rho/\sigma)$ & $ \Omega_{\Lambda}$ & {$H {\cal{T}}$} &$w$&

\multicolumn{1}{|c|}{$|g|$} \\ \hline\hline
                    $- 0.536$& $- 0.707$ & $0.031$& $0.695$&$1.843$& $- 0.691$&$0.066$   \\ \hline
$- 0.543$ & $- 0.703$  & $0.015$ &$0.698$&$1.932$ &$- 0.695$&$0.033$  \\ \hline
$- 0.550$ & $- 0.700$  & $0.000$ &$0.700$&$2.222$ &$- 0.700$&$0.000$  \\ \hline

\end{tabular}
\end{center}

\medskip

Turning back to the formulae in terms of the redshift $z$, we can use (\ref{numerical values}) to obtain the constants in (\ref{cosmological parameters  as a function of z}). In particular, for $\bar{\eta}$ we obtain 
\begin{equation}
\bar{\eta} = k_{(5)}^2 \sigma_{0}(\beta + 1)\bar{t} \left[ 4 + k_{(5)}^2 \sigma_{0}(\beta + 1)\bar{t}\right] =  22.7715,
\end{equation}
and $(\bar{\rho}/\sigma_{0}) = 0.02596$.  With these values, we can now use (\ref{cosmological parameters  as a function of z}) to reconstruct the past of the universe.  This model has a number of interesting properties but we leave their discussion to another opportunity.

\subsection{The future}
For $t \gg t_{0}$, far from the initial singularity, $\sigma \approx \sigma_{0}$ and $\sigma \gg \rho$. The asymptotic evolution of the expansion scale factor, as  $t \rightarrow \infty$,  is given by
\begin{equation}
\label{asymptotic scale factor}
a(t) = const \times t^{2/3(\beta  + 1)}, 
\end{equation}
irrespective of the equation of state of the matter on the brane. Thus, for the future evolution of 
the universe, we obtain  a cosmological model with constant $G$ and a {\em non-vanishing} cosmological term $\Lambda_{(4)}$, which varies with time. Namely,
\begin{equation}
\label{effective Lambda 4.1}
8\pi G \rightarrow k_{(5)}^4 \frac{\sigma_{0}}{6}, \;\;\;  \Lambda_{(4)} \rightarrow \frac{3(\gamma - \beta)}{(\gamma + 1)}H^2, \;\;\; H \rightarrow \frac{2}{3(\beta + 1)} \frac{1}{t} .
\end{equation}
The matter density and pressure become
\begin{equation}
\label{asymptotic density}
8\pi G \rho \rightarrow \frac{4}{3(\gamma + 1)(\beta + 1)t^2}, \;\;\;\;\; p = \gamma \rho.
\end{equation}
Here $\beta$ is related to the deceleration parameter as $ \beta \approx (2q - 1)/3$. We also note that the ratio $\dot{G}/G$ varies from $(\dot{G}/G) \approx - 3(\beta + 1)$ for $t \approx 0$ to $(\dot{G}/G) \rightarrow 0$ for $t \rightarrow \infty$.

 At this stage,   the parameter  $\beta$ becomes totally determined by $\Omega_{\Lambda}$. Namely, for any given $\gamma$ 
\begin{equation}
\label{beta through Omega Lambda}
\beta = - \Omega_{\Lambda} = w.
\end{equation}
The corresponding effective, or total,  energy density is obtained from (\ref{effective Lambda 4.1}) and (\ref{asymptotic density}) as follows
\begin{equation}
\label{General effective density}
8 \pi G \rho_{t} = \Lambda_{(4)} + 8 \pi G\rho = \frac{4}{3(\beta + 1)^2 t^2}, \;\;\;p_{t} =  - \Omega_{\Lambda} \rho_{t}.
\end{equation}
In this limit $\Omega_{\rho} + \Omega_{\Lambda} = 1$, which means that the brane-model under consideration grows up to become,  asymptotically in time,  indistinguishable from  a spatially flat FRW cosmological model based on a mixture of cold dark matter and a variable cosmological ``constant". The model can be completely determined by measurements of only one cosmological parameter. For example, if we know $\Omega_{\rho}$, then from (\ref{q}) and (\ref{equation of state for the whole dust universe}) we get $q$ and $w$, respectively. This in turn gives $\beta$ from (\ref{beta through Omega Lambda}), which allows to calculate the age of the universe and the Hubble parameter. 

\section{Summary and conclusions}

An important  
feature of braneworld theories is that the gravitational coupling  $G$ and the effective cosmological term $\Lambda_{(4)}$ are related to $\sigma$, the vacuum energy density 
  of the $3$-brane. They are {\em not} independent, 
 as in Jordan-Brans-Dicke and other multidimensional theories \cite{Kirill}-\cite{Melnikov}. Thus, in brane-theory  either  all of them are constants or they vary simultaneously.

In the case where the  vacuum energy is constant, and the ordinary matter satisfies the equation of state $p = \gamma \rho$,  the conservation equation $T^{\mu}_{\nu; \mu} = 0$ leads to $\rho \sim a^{-3(\gamma + 1)}$. 
In this paper, in order to integrate the generalized FRW equation of braneworld models, we have  assumed $\rho \sim a^{- 3(\beta + 1)}$, where $\beta \neq \gamma$ is a constant. We have seen that $\beta \approx \gamma$ at the early universe, while  $\beta \approx - \Omega_{\Lambda}$ at late times.
This assumption  constitutes the ``minimum" extension to the usual braneworld models with constant vacuum energy $(\beta = \gamma)$. It adds new algebraic possibilities, while keeping the  calculations simple enough as to lead to physical effects that can be studied exactly.

The expansion of the universe at late times is determined by  the constant $\Lambda_{eff}$ introduced in (\ref{eff Lambda}). We have set $\Lambda_{eff} = 0$, thus excluding from the outset the existence of recollapsing or exponentially expanding solutions, and 
studied in detail the spatially-flat cosmological model with  ${\cal{C}} = 0$. 

In general $\beta$ is a ``function" of the cosmological parameters $q$ and $\Omega_{\rho}$. If $\beta = \gamma$, then from (\ref{quadratic equation}) we obtain $q = 2 - 3\Omega_{\rho}/2$,  in the dust dominated era. Thus fixing $\Omega_{\rho}$ today also fixes $q$. Notice that $q$ is positive for any physical value of $\Omega_{\rho}$, meaning that a brane-universe with constant vacuum energy must be slowing down its expansion. However, for $\beta \neq \gamma$, this is no longer so; the cosmological parameter $\Omega_{\Lambda}$ is now a dynamical quantity which changes this picture. The acceleration parameter is now given by (\ref{q}) and (\ref{q and w as functions of gamma}), so that as the universe expands $q$ decreases from $q_{\gamma} = 3\gamma + 2$ at the beginning of the universe to $q \rightarrow - 1 + 3\Omega_{\rho}/2$ at late times. Thus, leading to a phase where the universe, instead of slowing down,  is speeding up its expansion.    

As a consequence of the introduction of $\beta \neq \gamma$, the cosmological  parameters $q$ and $\Omega_{\rho}$ are independent from each other. However, for any given value of $\Omega_{\rho}$ (or $q$), the positivity of energy and the observational limits on the variation of $G$,  allow to determine $q$ (or $\Omega_{\rho}$) with a precision which is close to $100 \%$. 

Therefore, the equation of state of the universe (\ref{equation of state for the whole dust universe}) varies very slowly in each epoch of the universe (see table $6$). But it does change when $\gamma$ changes, from  (\ref{w}) it follows that $w$ is affected by $\gamma$ through its influence on $q$.  As the universe expands the equation of state changes from  $w \approx 1 + 2\gamma$ at the very early universe, to become quintessence-like at late times, namely, $w \rightarrow - 1 + \Omega_{\rho}$. Recent  data from type Ia supernovae in distant galaxies and the cosmic microwave background   
constrain the equation of state of the universe \cite{Efstathiou}. If the universe is assumed to be spatially flat, as we do here, then the upper limit for $w$ is $- 0.6$, which is close to the lower limit $w \approx - 0.7$ allowed for  quintessence tracker fields 
 \cite{Steinhard}. These limits\footnote{Since $\Omega_{\rho} > 0$, our braneworld  model excludes phantom energy, which requires $w < - 1$} are quite similar to those derived  here (\ref{specific number for w}).

A remarkable feature of our model is that $\Omega_{\rho}$ is bounded above. Consequently, at some stage of the expansion the vacuum energy starts dominating the evolution. Precisely, the density parameter reaches its maximum value when the epoch where $\rho > \sigma$ comes to an end. From then on $\sigma > \rho$. One might think that at this stage something radical happens. In particular, that the evolution becomes vacuum dominated $(\Omega_{\Lambda} > \Omega_{\rho})$. However, because of the quadratic correction which is still considerable at this stage, this is not so. The reality is that for most of the evolution of the universe $\Omega_{\rho} > \Omega_{\Lambda}$. Only recently, for $H{\cal{T}}\approx 0.8$ (see table $5$), the black energy overtakes  the matter energy and drives the universe into a period of accelerated expansion. 

The model discussed here grows to be  FRW-like in time, in the sense that the quadratic correction $\Omega_{\rho^2}$ becomes negligible and $G$ becomes constant at late times. However, there is an important  distinction; the cosmological ``constant" varies with time and is proportional to $H^2$. It should be noted that the relation  $\Lambda_{(4)} \approx const \times H^2$ is favored by observations and there is an extensive literature suggesting that it plays a fundamental role in cosmology \cite{Overduin}-\cite{Vishwakarma}.

Another important feature of the model is that it contains no ``adjustable" parameters. All the quantities, even the five-dimensional ones, can be evaluated  by means of  measurements  in $4D$. We illustrate  the values for $k_{(5)}$ and $\Lambda_{(5)}$ in (\ref{5D quantities}) and (\ref{numerical values}). We remark that these are ``universal" constants fixed by the five-dimensional embedding bulk. Regarding the ``cosmic coincidence" problem \cite{Zlatev}, it is interesting to note from (\ref{numerical values}),  that in our model 
the missing energy density $\bar{\rho}_{\Lambda}$ is close to $\bar{\rho}$, the matter density today, without any (obvious) fine-tuning of their ratio at the early universe. 

Let us briefly compare this work with our previous investigation \cite{jpdel2}. In that paper our working hypothesis was that $(\dot{G}/HG) = g $, remained constant throughout  the evolution of the universe.  Then, the observational constraint $|g| < 0.1$ required $\sigma_{0} = 0$. This means that, as the universe expands $\sigma \rightarrow 0$ and, from (\ref{definition of lambda}),  $\Lambda_{(4)} \rightarrow k_{(5)}^2 \Lambda_{(5)}/2$.  If our universe  is assumed to be  embedded in an anti-de Sitter five-dimensional bulk, then $\Lambda_{(4)}$ changes its sign, from positive at the beginning to negative later, which in turn causes the universe to recollapse in the remote future $(H{\cal{T}} \approx 10.76)$. For a universe embedded in a  bulk with  $\Lambda_{(5)} = 0$, we obtained the ever-expanding solution given by (\ref{Case already discussed}) and (\ref{previous investigation}). 
In the present work $\sigma_{0} \neq 0$ and there is neither recollapse, nor exponential expansion. We excluded both by setting $\Lambda_{eff} = 0$. 

In spite of this, and the fact that the recollapsing model is few times older than  the present one, the cosmological parameters $q$, $\Omega_{\Lambda}$, $\Omega_{\rho^2}$ as well as the five-dimensional quantities $k_{(5)}$ and $\Lambda_{(5)}$ have similar values in both models.  
Therefore, we cannot discriminate between them on the basis of the cosmological parameters only. In order to do that, another type of analysis should be done, for example the behavior of perturbations and structure formation. These problems are beyond the scope of the present work.

In summary, in this paper we have constructed a cosmological model, in the context of five-dimensional braneworld theory, where the vacuum energy varies with time. The model provides a simple scenario where we can study the simultaneous variation of $G$ and $\Lambda_{(4)}$  in an universe which is spatially flat and expanding with acceleration, in agreement with recent observations. The vacuum density $\sigma$ may vary in models where $g_{44} = - \Phi^2$ is a function of time. In this connection, it is important to mention that the ratio $(\dot{\Phi}/\Phi)$ appears  in different contexts, notably in expressions concerning the variation of rest mass and electric charge. Regarding the time-variation of the gravitational ``constant" $G$, in different models with extra dimensions \cite {Melnikov} the ratio $(\dot{G}/G)$ is found to be proportional to $(\dot{\Phi}/\Phi)$. 

Therefore, we have a scenario where  the observed acceleration of the universe is just one piece in the dynamical evolution of an universe where the so-called fundamental ``constants" are evolving in time. The fact that  asymptotically in time $G$ becomes constant, suggests that the size of the extra dimension stabilizes in time  and, consequently, the rest mass and the electric charge become constant too.  
This study can be extended by considering other possible functional forms for the vacuum energy, obtained from realistic potentials $V(\phi)$ determined from the observed cosmological functions. 
\renewcommand{\theequation}{A-\arabic{equation}}
  \setcounter{equation}{0}  
  \section*{Appendix: The vacuum energy as the energy of a slowly-evolving scalar field}  

Although the energy-momentum tensor for vacuum looks like an ideal fluid with negative pressure, the dynamics of a varying vacuum cannot be that of a fluid because stability would require $dp/d\rho >0$.  The simplest working model for a dynamical vacuum energy is the dark energy of a single homogeneous  scalar field $\phi$ (its slight clumping is neglected) with self-interaction potential $V(\phi)$. 

For completeness, we would like to show here that the vacuum energy density (\ref{variable density}) can be derived from a simple potential.

For the cosmological metric (\ref{cosmological metric}) the equation governing $\phi$ is 
\begin{equation}
\label{field equation}
\ddot{\phi} + 3 \frac{\dot{a}}{a}\dot{\phi} + \frac{d V}{d \phi} = 0.
\end{equation}
The energy density $\rho_{\phi}$ and pressure $p_{\phi}$ are given by 
\begin{eqnarray}
\rho_{\phi} &=& \frac{1}{2}{\dot{\phi}}^2 + V(\phi),\nonumber \\
p_{\phi} &=& \frac{1}{2}{\dot{\phi}}^2 - V(\phi).
\end{eqnarray}
Here the appropriate dimensions  for  $\phi$ are $cm^{- 3/2}gr^{1/2}s^1$. If $V(\phi)$ is sufficiently flat, then $\phi$ will evolve very slowly and ${\dot{\phi}}^2 \ll V$. In this scenario the field energy approximates the effect of a cosmological term with 
$p_{\phi} \approx - \rho_{\phi}$. 

Combining the above equations we get
\begin{equation}
{\dot{\phi}}^2 = - \frac{a}{3}\frac{d\rho_{\phi}}{da}.
\end{equation}
Consequently,
\begin{equation}
V = \rho_{\phi} + \frac{a}{6}\frac{d\rho_{\phi}}{da}.
\end{equation}
We now identify the vacuum energy with the energy of the scalar field, viz., $\sigma = \rho_{\phi}$. Then using (\ref {variable density}) we obtain
\begin{eqnarray}
V &=& \sigma_{0} + \frac{(\gamma - \beta)(1 - \beta)}{2(\beta + 1)}\rho,\nonumber \\
{\dot{\phi}}^2 &=& {(\gamma - \beta)}\rho.
\end{eqnarray}
For the solution under consideration (\ref{evolution for vanishing lambda eff}) we get
\begin{equation}
\label{potential for our solution}
V(\phi) = \sigma_{0} + \frac{(\gamma - \beta)(1 - \beta)\sigma_{0}}{(\gamma + 1)\sinh^2{C (\phi - \phi_{0}})},
\end{equation}
where $\phi_{0}$ is a constant of integration and\footnote{For the evaluation of $C$ we have used the numerical values found in (\ref{numerical values}).} 
\begin{equation}
C \equiv \sqrt{\frac{k_{(5)}^4\sigma_{0}(\beta + 1)(\gamma + 1)}{8(\gamma - \beta)}}
 \approx 0.704 \times 10^{- 3} cm^{3/2}gr^{- 1/2}s^{- 1}.
\end{equation}
The time evolution of the scalar field is given by
\begin{equation}
C(\phi - \phi_{0}) = \ln\left[\frac{(\tau + 2) + \sqrt{\tau^2 + 4 \tau}}{2}\right].
\end{equation}
We notice that $V(\phi)$ sharply decreases with the increase of $\phi$, so that $V(\phi) \approx \sigma_{0}$ and $p_{\phi} \approx - \rho_{\phi}$ for most of the evolution of the universe. Potential (\ref{potential for our solution}) takes a more familiar form in the early universe and the late universe.

\paragraph{Power law expansion} In an early universe, dominated by matter or radiation, the term $\rho^2$ drives power law expansion $a \sim t^{1/3(1 + \beta)}$. At late times the dark energy becomes dominant and also drives a power law expansion, viz.,  $a \sim t^{2/3(1 + \beta)}$. In general, for a scale factor $a \sim t^n$ the field equation (\ref{field equation}) has two types of solutions. The first one corresponds to the pure inverse power-law potential, namely,
\begin{equation}
\label{first solution}
V(\phi) = \sigma_{0} + \frac{const}{\phi^{\alpha}}, \;\;\; \phi \sim t^{2/(2 + \alpha)},
\end{equation}
where $\alpha$ is a dimensionless parameter related to the equation of state of the universe. 
 The second solution is for the exponential potential  
\begin{equation}
\label{second solution}
V(\phi) = \sigma _{0} + \frac{4(1 - \beta)(\gamma - \beta)\sigma_{0}}{(\gamma + 1)}e^{- 2C\phi}, \;\;\;\; C(\phi - \phi_{0}) =  \ln \tau. 
\end{equation}
In our solution, the potential (\ref{potential for our solution}) can be expanded in powers of $\phi$. The first term of the expansion  corresponds to (\ref{first solution}) with $\alpha = 2$. On the other hand (\ref{second solution}) is the asymptotic $(\tau \rightarrow \infty)$ form  of (\ref{potential for our solution}). 

Therefore, for the largest part of the evolution the scalar potential  (\ref{potential for our solution}) is a good model for the cosmological term. But, notice that it does not work at the very early universe where $p_{\phi} \approx - \beta \rho_{\phi} \approx - \gamma \rho_{\phi}$; another mechanism for a variable  cosmological term should be used for that era.

Before finishing, we would like to mention an important difference\footnote{Besides the fact that $G$ is constant in FRW models and here it should vary with time.} between the present model and the FRW cosmologies with  a variable cosmological term, modeled as a  scalar field. Namely, in FRW models $\Lambda_{(4)}$ is a linear function of the vacuum energy, viz., $\Lambda_{(4)} \sim \rho_{\phi}$, while here it is a quadratic function; $\Lambda_{(4)} = k_{(5)}^4 (\sigma^2- \sigma_{0}^2)/12$, with $\sigma = \rho_{\phi}$. In spite of this, the numerical values for the vacuum energy and cosmological constant as calculated in (\ref{numerical values}) are similar to those expected from other branches of physics.

\end{document}